\documentclass[preprint2]{aastex}

\usepackage{ulem}
\usepackage{color}

\newcommand{\Msun}{M_\odot}
\newcommand{\sub}[1]{_{\rm #1}}
\newcommand{\amin}{a_{\rm crit}}
\newcommand{\SPGP}{SPGPs}

\slugcomment{}
\shorttitle{Planet Engulfment by $\sim$1.5--3 $\Msun$ Red Giants}
\shortauthors{Kunitomo et al.}

\begin{document}

\title{Planet Engulfment by $\sim$1.5--3 $\Msun$ Red Giants}

\author{M. Kunitomo\altaffilmark{1}, 
M. Ikoma\altaffilmark{1}\altaffilmark{*}, 
B. Sato\altaffilmark{1, 2}, 
Y. Katsuta\altaffilmark{3}, and 
S. Ida\altaffilmark{1}
}


\altaffiltext{1}{Department of Earth and Planetary Sciences, 
Tokyo Institute of Technology (TokyoTech), 2-12-1 Ookayama, Meguro-ku, 
Tokyo 152-8551, Japan}
\altaffiltext{2}{Global Edge Institute, Tokyo Institute of Technology (Tokyo Tech), 
2-12-1 Ookayama, Meguro-ku, Tokyo 152-8551, Japan}
\altaffiltext{3}{Department of Cosmosciences, 
Hokkaido University, Kita 10 Nishi 8, Kita-ku, Sapporo 060-0810, Japan}
\altaffiltext{*}{Corresponding author: M. Ikoma (mikoma@geo.titech.ac.jp)}

\begin{abstract}
Recent radial-velocity surveys for GK clump giants have revealed that planets also exist around $\sim$1.5--3 $\Msun$ stars. However, no planets have been found inside 0.6~AU around clump giants, in contrast to solar-type main-sequence stars, many of which harbor short-period planets such as hot Jupiters.
In this study we examine the possibility that planets were engulfed by host stars evolving on the red-giant branch (RGB). 
We integrate the orbital evolution of planets in the RGB and helium burning (HeB) phases of host stars, including the effects of stellar tide and stellar mass loss. Then we derive the critical semimajor axis (or the survival limit) inside which planets are eventually engulfed by their host stars after tidal decay of their orbits.  
Especially, we investigate the impact of stellar mass and other stellar parameters on the survival limit in more detail than previous studies. In addition, we make detailed comparison with measured semimajor axes of planets detected so far, which no previous study did.
We find that the critical semimajor axis is quite sensitive to stellar mass in the range between 1.7 and 2.1 $\Msun$, which suggests a need for careful comparison between theoretical and observational limits of existence of planets. Our comparison demonstrates that all those planets are beyond the survival limit, which is consistent with the planet-engulfment hypothesis. However, on the high-mass side ($> 2.1 \Msun$), the detected planets are orbiting significantly far from the survival limit, which suggests that engulfment by host stars may not be the main reason for the observed lack of short-period giant planets.
To confirm our conclusion, the detection of more planets around clump giants, especially with masses $\gtrsim 2.5 \Msun$, is required.
\end{abstract}


\section{INTRODUCTION \label{sec: introduction}}
Detection of a significant number of exoplanets orbiting Sun-like stars (FGK dwarfs) has helped us to improve our understanding of the origins and diversity of planetary systems.  Statistics on detected planets suggests the presence of several different populations and correlations.  That has motivated theorists to synthesize planet populations by integrating the formation process from planetesimals to planets \citep[][]{Ida+Lin04, Ida+Lin04b, Ida+Lin05, Ida+Lin08a,Ida+Lin08b,Ida+Lin10,Kennedy+Kenyon08a,Kennedy+Kenyon08b,Mordasini+09a,Mordasini+09b}.  Comparison with statistical properties from observations has then enabled us to calibrate theories for planet formation.

As for heavier stars such as BA dwarfs, the situation is somewhat different.  Precise radial-velocity measurement is not successful, because of the lack in absorption lines on their hot photospheric surfaces. Instead, efforts have been made to find planets around GK giants that used to be BA dwarfs on their main-sequence. Thus, comparison between observations and theories is not straightforward, unlike in the case of FGK dwarfs. That means we have to take into account modifications to orbital configurations of planets during their host stars' evolution. 

According to recent radial-velocity surveys for GK clump giants, it appears that there is a lack of giant planets inside 0.6~AU \citep[][and references therein]{Sato+10}. Such a deficit is not seen around Sun-like stars which often harbor short-period planets such as hot Jupiters.  It is unclear whether the deficit is primordial---namely, short-period giant planets ({\SPGP}) are not formed around BA dwarfs originally; otherwise short-period planets are removed during host stars' evolution. The latter possibility is focused on in this paper.

A star evolves off the main sequence (MS) towards the red-giant branch (RGB), after exhaustion of hydrogen at its center. Then, once the helium ignites, the star enters a next central nuclear burning phase that is sometimes called the helium-burning phase (HeB). Most of the clump giants with detected planets are thought to be on their HeB. Before reaching the HeB, namely, in the RGB phase, the star expands substantially, becomes highly luminous, and may lose substantial mass, which should affect orbits of surrounding planets.

The evolution and fate of planets in the RGB phase of their host stars have been investigated in the context of the survival of the Earth. Based on his simple estimation, \citet{Vila84}, followed by a somewhat detailed argument by \citet{Goldstein87}, showed that the Earth would lose its momentum quickly in the extending envelope of the Sun before the envelope starts to shrink again. That means the Earth would be unable to escape from the Sun once engulfed. Those studies suggest the need for the detailed simulation of the Sun's evolution. The Sun's evolution, including solar mass loss, was simulated in detail by \citet{Sackmann+93}, who demonstrated that the decrease in the solar mass is effective in pushing the terrestrial planets other than Mercury outwards and preventing them from being engulfed by the expanded Sun. Later, building on theories for formation of close binaries \citep[e.g.,][]{Livio82, Livio+Soker83, Livio+Soker84}, \citet{Rasio+96} incorporated the tide raised on the Sun and then demonstrated that Venus is likely to fall into the Sun because of the solar tide enhanced by the expansion of the envelope, while the Earth's fate remained inconclusive, depending on unconstrained parameters for tide. Other effects such as the gravitational drag, ram pressure, wind-accretion \citep{Vila84, Duncan+Lissauer98}, and orbital instability due to secular perturbation \citep{Duncan+Lissauer98} are known to be negligible. Note that the last possibility could be important in extrasolar planetary systems, depending on orbital configurations.

To understand the observed lack of {\SPGP} around GK clump giants, \citet{Sato+08} and \citet{Villaver+Livio09} investigated the orbital evolution of planets around intermediate-mass RGB stars in a similar way to \citet{Rasio+96}.
They demonstrated that 
{\SPGP} did undergo orbital decay due to stellar tide, ending up being swallowed by their host stars. {In particular, \citet{Villaver+Livio09} derived a critical semimajor axis, $\amin$ (note that they called it a minimum semimajor axis and denoted it by $a\sub{min}$), beyond which planets evade engulfment by their host stars during the RGB phase: In four cases of different stellar initial masses, $M_{\star, i}$ (i.e., 1, 2, 3, and 5~$\Msun$), the values of $\amin$ that they derived  for Jovian-mass planets were 3.0, 2.1, 0.18, and 0.45~AU, respectively. Then, without detailed comparison of $\amin$ with the orbital distribution of known planets around GK clump giants,} they concluded that engulfment by host stars may be responsible for the observed lack of {\SPGP} inside 0.6~AU and any ad hoc mechanism to never make {\SPGP} may not be required.

{In this paper, to verify the planet-engulfment hypothesis, we make a detailed comparison of the theoretical limit of survival (namely, $\amin$) with semimajor axes of the planets detected so far around GK clump giants.}
{The mass of those giants ranges from $\sim 1.5$ and $3$~$\Msun$.} As known in the study of stellar evolution \citep[e.g.,][]{Kippenhahn+Weigert90}, the RGB-tip radius is sensitive to stellar mass in such a range. 
Given that stellar tide depends strongly on stellar radius (see section \ref{sec:method}), the limit of survival from engulfment by host stars should be sensitive to host stars' mass, which implies the need for detailed comparison between the theoretical and observational limits. {It should be noted that \citet{Nordhaus+10} has recently made a thorough study of the tidal evolution of orbits of planets around post-MS stars. Their focus was, however, on planets around white dwarfs that survive engulfment by AGB stars. Of interest in this paper is planets around RGB or HeB stars that have not entered the AGB phase yet.}

This paper is organized as follows. In section~\ref{sec:method}, we first describe our physical model and computation method to simulate the orbital evolution of a planet around its host star that is evolving off the MS. In section~\ref{sec:results}, we derive a critical semimajor axis beyond which planets survive the host stars' RGB/HeB phase, and investigate its sensitivity to stellar mass and other parameters. 
In section~\ref{sec:discussion}, we compare the derived survival limit with measured semimajor axes of the detected planets. {The comparison suggests that planet engulfment may not be the main reason for the observed paucity of {\SPGP}, apart from statistical sufficiency.}
{In section~\ref{sec:validity}, we evaluate the impacts of several uncertainties on the survival limit and find that our conclusion is not affected by them.} 
{In section~\ref{sec:primordial}, we mention the possibility of stellar-mass-dependent formation process that may account for the observed paucity of {\SPGP}.}
{Finally, we summarize this paper, claiming the need for more planet samples around giants of $> 2.5 \Msun$ to confirm the validity of our findings, in section~\ref{sec:conclusion}.}

\section{PHYSICAL MODEL AND COMPUTATION METHOD \label{sec:method}}

We simulate the evolution of orbits of planets during host stars' evolution. {We assume circular orbits, because our focus is on semimajor axis in this study. 
The eccentricities of planets detected around the GK clump giants that we target in this study are as small as 0.25. Such a small eccentricity has a small effect on the evolution of semimajor axis. The impact of the finite eccentricity on the limit of survival is evaluated in section~\ref{sec:e}. Note that previous studies \citep[][]{Villaver+Livio09, Nordhaus+10} also assumed zero eccentriticies. } 

We consider the effects of stellar tide and mass loss on the planetary orbit. We neglect other competing processes such as the frictional and gravitational drag forces by stellar wind and change in the planet mass due to stellar-wind accretion and evaporation, which were evaluated to be negligible by \citet{Villaver+Livio09} and \citet{Duncan+Lissauer98}. Thus, we
integrate the equation, 
   \begin{equation}
     \frac{1}{a} \frac{\mathrm{d}a}{\mathrm{d}t} = 
        -6 \frac{k}{T} \frac{M_p}{M_\star} 
	               \left(1+\frac{M_p}{M_\star}\right)
		       \left(\frac{R_\star}{a}\right)^8
	-\frac{\dot{M}_\star}{M_\star},
   \label{eq:orbit}
   \end{equation}
from the zero-age main-sequence of the host star of a given mass. In the above equation, $a$ is the semimajor axis, $t$ is time, $M_p$ is the planet's mass, $M_\star$ and $R_\star$ are the mass and radius of the host star, respectively, $k$ is the apsidal motion constant, and $T$ is the eddy turnover timescale (see below). The first term on the right-hand side represents the effect of stellar tide \citep{Hut81}. Since RGB stars are slow rotators, we assume no stellar rotation, which means that the stellar tide always causes orbital decay of the planet.

As for the parameters for stellar tide such as $k$ and $T$, following \citet{Rasio+96} and \citet{Villaver+Livio09}, we adopt the turbulent viscosity: 
   \begin{equation}
     k = \frac{f}{6} \frac{M\sub{env}}{M_\star},
     \label{eq:k}
   \end{equation}
where $M\sub{env}$ is the mass in the convective envelope and $f$ is given by 
   \begin{equation}
     f = \min \left[1, \left(\frac{P}{2T}\right)^2 \right]
     \label{eq:f}
   \end{equation}
with $P$ being the orbital period.  The eddy turnover timescale $T$ is given by 
   \begin{equation}
     T = \left[\frac{M\sub{env}(R_\star-R\sub{env})^2}{3L_\star}\right]^{1/3},
     \label{eq:T}
   \end{equation}
where $R\sub{env}$ is the radius at the base of the convective zone and $L_\star$ is the stellar intrinsic luminosity. {
The factor $f$ is introduced to weaken the stellar tide when the orbital period is shorter than the eddy turnover timescale. However, it is a matter of debate how to deal with such fast tide \citep[e.g.,][]{Zahn08}. We discuss the impact of the uncertainties in $f$ on the survival limit in section~\ref{sec:tidal model}.}

The second term on the right-hand side of equation~(\ref{eq:orbit}) represents orbital migration due to stellar mass loss ($\dot{M}_\star < 0$). We use the Reimers' parameterisation for stellar mass loss
\citep{Reimers75}, namely, 
   \begin{equation}
     \dot{M}_\star = - 4 \times 10^{-13} \eta 
     			\left(\frac{L_\star}{L_\odot}\right)^{}
     			\left(\frac{R_\star}{R_\odot}\right)^{}
     			\left(\frac{M_\star}{\Msun}\right)^{-1} 
			\Msun \, {\rm yr}^{-1},
   \end{equation}
where $\eta$ is the mass-loss parameter of order unity.

In this study, we simulate stellar evolution directly with the code MESA ver.~2258  
\citep[][]{Paxton+10} to calculate $M_\star (t)$, $R_\star (t)$, $L_\star (t)$, $\dot{M}_\star (t)$, $M\sub{env} (t)$, and $R\sub{env} (t)$. 
We adopt the mass-loss parameter $\eta = 0.6$ and the mixing-length parameter $\alpha=1.5$. We include the effects of convective overshooting, following the MESA's prescription where the overshoot parameters are set to be 0.0128 at the bottom of non-burn regions and 0.014 at all the other boundaries. {Several different stellar-evolution models that are simulated with different numerical codes are available in the literature and online. \citet{Villaver+Livio09} used a stellar-evolution code other than MESA. We discuss the differences and their impacts on the survival limit in section~\ref{sec:code}.}

\section{THEORETICAL LIMIT OF SURVIVAL \label{sec:results}}
\subsection{Orbital Evolution \label{sec: orbital evolution}}
\begin{figure}[tb]
  \begin{center}
    \includegraphics[width=8cm,keepaspectratio]{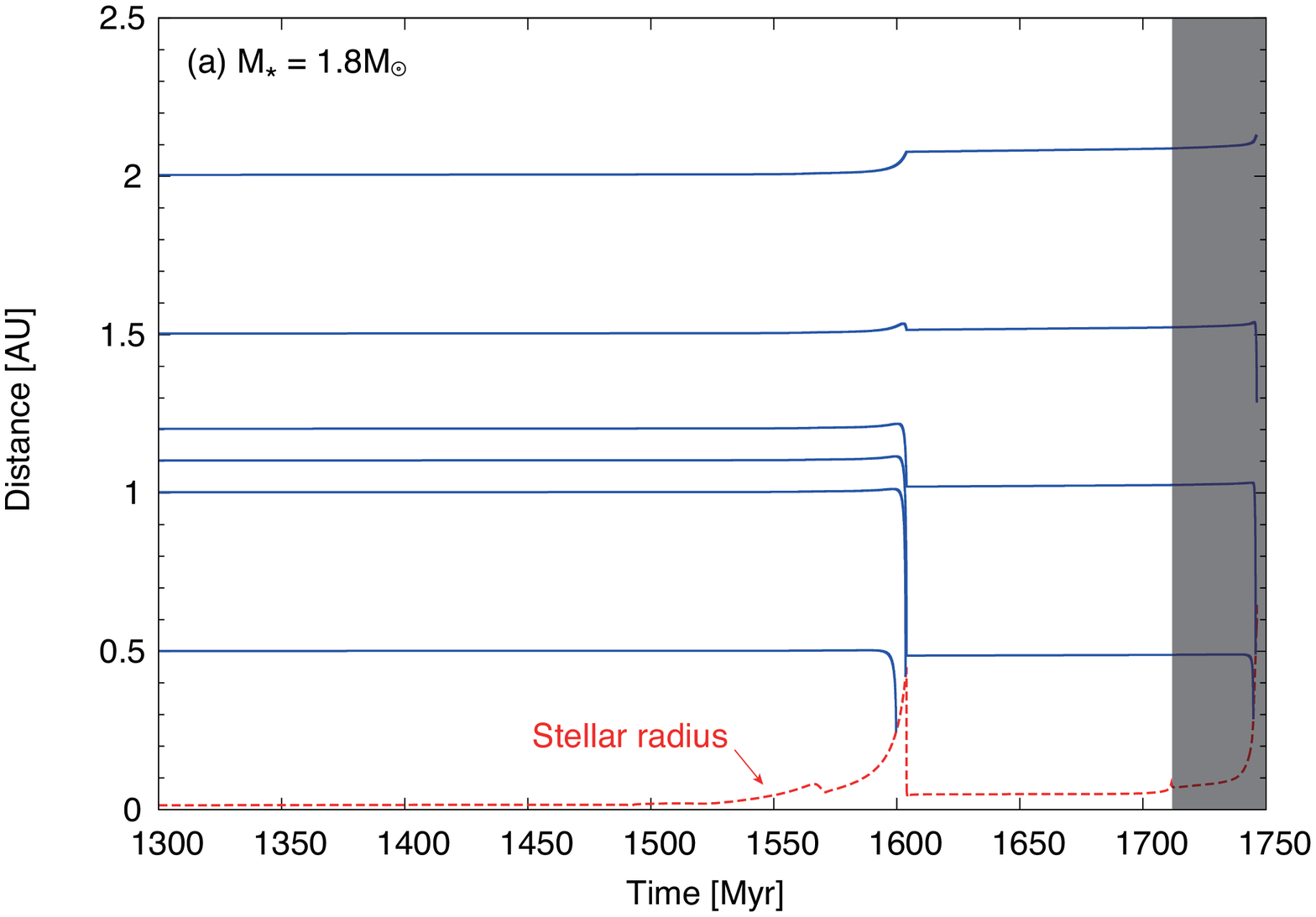}
    \includegraphics[width=8cm,keepaspectratio]{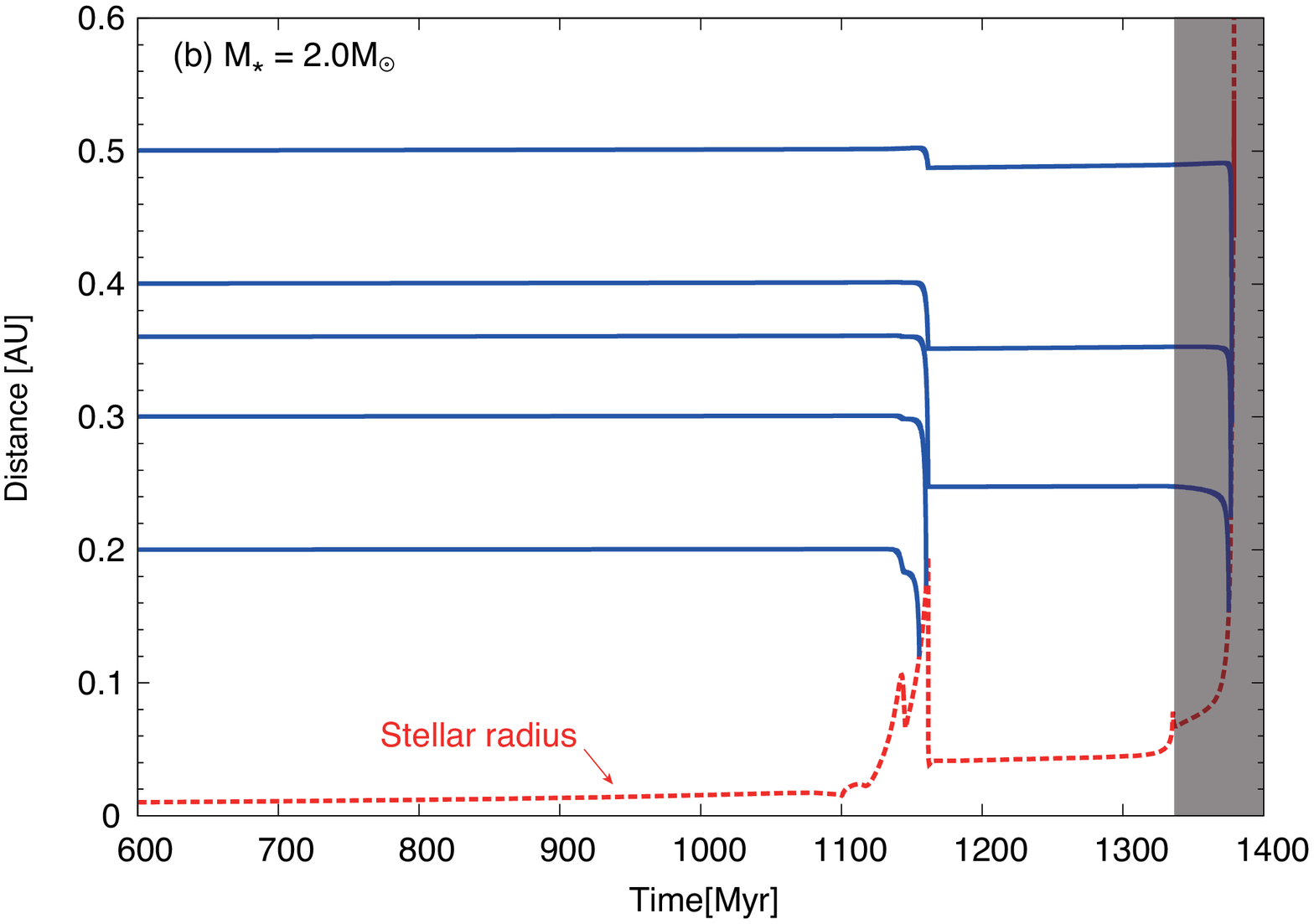}
    \caption{\small{Orbital evolution of 1-Jupiter-mass planets with different initial semimajor axes (solid lines) for $M_\ast =(a) $~1.8~$\Msun$ and (b) 2.0~$\Msun$; the metallicity, $Z_\star$, is 0.02 in both cases. Dashed lines represent the evolution of stellar radii.}  \label{fig:orbital evolution}}
  \end{center}
\end{figure}

We derive the theoretical limit of survival from engulfment by evolving host stars. Figure~\ref{fig:orbital evolution} shows examples of the orbital evolution of planets around stars with initial masses, $M_{\star, i}$, of (a) $1.8 \Msun$ and (b) $2.0 \Msun$; the metallicity is 0.02. We have integrated the orbit of a 1$M\sub{J}$ planet ($M\sub{J}$: Jupiter's mass) with different initial semimajor axes (solid lines) from the host-star's zero-age main sequence until the planet's semimajor axis falls below the stellar radius ($a < R_\star$) or until the AGB thermal pulse (AGB-TP) occurs. The evolution of the stellar radii is represented by the dashed lines.

{Although we have continued the simulations after the end of HeB (grey areas in the figure), 
our interest in this study is in whether planets survive the RGB and HeB phases of their host stars. 
This is because the planet-harboring GK clump giants that we target for the comparison of our survival limit with the semimajor axes of their planets are thought to be in the HeB or RGB phases, as stated in Introduction.} 
{As for orbital evolution of planets around AGB stars, \citet{Nordhaus+10} made a thorough investigation.}

As seen in Fig.~\ref{fig:orbital evolution}a, the 1.8$\Msun$ star expands up to $0.5$~AU at the RGB-tip. As the stellar radius becomes large, tidal decay is rapidly enhanced (eq.~[\ref{eq:orbit}]), so that the star consequently swallows planets whose initial semimajor axes, $a_{i}$, are smaller than 1.1~AU.  The planet that starts out at 1.1~AU is tidally pulled by the host star, but it barely survives the RGB-tip and remains in the HeB phases. Outer planets (e.g., $a_{i} = 2.0$~AU) are found to be pushed outwards because of stellar mass loss; the effect is, however, small, unlike in the case of 1$\Msun$ stars, as also noted previously \citep{Sato+08, Villaver+Livio09, Nordhaus+10}. As for the 2$\Msun$ star, the RGB-tip radius is as small as 0.20~AU (Fig.~\ref{fig:orbital evolution}b).  As a consequence, even a planet starting out at 0.36~AU can avoid engulfment during the RGB/HeB phases. 

We define a critical initial semimajor axis ($\amin$) below which planets end up being engulfed by their host stars at some point on the RGB or HeB. For the above two examples, $\amin = 1.1$~AU around the 1.8$\Msun$ star, while $\amin =$~0.36~AU around the 2$\Msun$ star.  It is thus illustrated that the survival limit is rather sensitive to the host star's mass. Furthermore, comparing Fig.~\ref{fig:orbital evolution}b with the upper-right panel of Fig.~1 of \citet{Villaver+Livio09}, one notices that the value of $\amin$ derived here for the 2$\Msun$ star is considerably different from that derived by \citet{Villaver+Livio09}, who estimated $\amin$ to be 2.1~AU, instead of 0.36~AU.  The reason for the difference is discussed in section~\ref{sec:code}.

For comparison with semimajor axes of the detected planets, the above definition of $\amin$ may be inappropriate, because it is the initial value of the semimajor axis of the planet that barely survives the host-star's RGB-tip. As shown in the above figure, the planet's semimajor axis after the RGB-tip is lower than its initial value. However, the probability that such planets are detected should be low. Because of unambiguity of definition, we decide to use the critical semimajor axis defined above.

\subsection{Critical Semimajor Axes \label{sec: survival limit}}
\begin{figure}[tb]
  \begin{center} 
    \includegraphics[width=8cm,keepaspectratio]{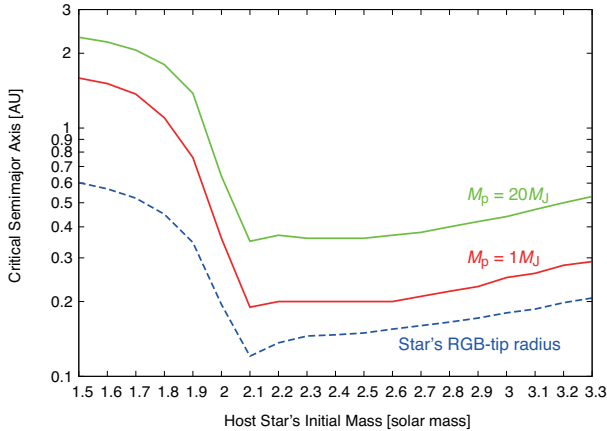}
    \caption{\small{The critical initial semimajor axis, $\amin$, inside which planets are engulfed by host stars before reaching the end of the helium-burning phase for two choices of planetary mass, $M_p$, namely, $M_p =$ 1~$M\sub{J}$ and 20~$M\sub{J}$. For comparison, the RGB-tip radius of the host star is shown by the dashed line. The stellar metallicity is 0.02 in the calculations.} \label{fig:amin vs. aorb}
    }
  \end{center}
\end{figure}

Figure~\ref{fig:amin vs. aorb} shows the relation between $\amin$ and $M_{\star, i}$  for $M_p = 1 M\sub{J}$ and 20~$M\sub{J}$. The stellar RGB-tip radius ($R\sub{tip}$) is also shown by the dashed line.

First, one notices that the curves for $\amin$ are similar in shape to that for $R\sub{tip}$. This is because the tidal decay rate of planetary orbit is quite sensitive to the stellar radius (see eq.~[\ref{eq:orbit}]). That means planet engulfment, in most cases, occurs right around the RGB-tip of the host star, at which the stellar radius takes a maximum during RGB/HeB phases. 
{To see that more clearly, we derive an approximate expression of $a\sub{crit}$ as follows.} From equation~(\ref{eq:orbit}), a typical decay timescale ($\tau\sub{decay}$) can be expressed by 
  \begin{equation}
  	\tau\sub{decay} \sim \frac{T}{6k} \frac{M_\star}{M_p} \left(\frac{R_\star}{a}\right)^{-8}, 
  \end{equation}
provided the mass-loss effect is negligible. If $\tau\sub{decay}$ is comparable to the typical duration of the RGB phase ($\tau\sub{RGB}$), then the planet is engulfed by its host star. Thus, $\amin$ would be expressed as
  \begin{eqnarray}
    \amin &=& \left(\frac{12 \pi^3 \sigma}{G} \right)^{1/5} 
      		      \frac{T\sub{eff}^{4/5} M_p^{1/5}}{M_\star^{3/5}} 
	               \tau\sub{RGB}^{1/5} R\sub{tip}^{8/5} \nonumber \\
	     &=& 0.11 
			     \left(\frac{R\sub{tip}}{0.1 \rm AU} \right)^{8/5}
	     		     \left(\frac{M_\star}{2 \Msun}\right)^{-3/5} 
			     \nonumber \\
		& &
	     		     \left(\frac{\tau\sub{RGB}}{5 {\rm Myr}}\right)^{1/5} 
			     \left(\frac{M_p}{M\sub{J}}\right)^{1/5}
			     \left(\frac{T\sub{eff}}{4000{\rm K}}\right)^{4/5}
			     {\rm AU} \nonumber \\
		& &
  \label{eq:acrit}
  \end{eqnarray}
In deriving this equation, based on our numerical results, we have assumed that $M\sub{env} \sim M_\star$ and $R\sub{env} \sim 0$ in equations~(\ref{eq:k}) and (\ref{eq:T}) and used $T \sim (M_\star R\sub{tip}^2 / 3 L\sub{tip})^{1/3}$ with the stellar RGB-tip luminosity $L\sub{tip} = 4 \pi \sigma R\sub{tip}^2 T\sub{eff, tip}^4$ and $k \sim$~$f$/6 $= (P/2T)^2/6$ (see eqs.~[\ref{eq:k}]-[\ref{eq:T}]).  Equation~(\ref{eq:acrit}) confirms that $R\sub{tip}$ is the most important factor and the others have minor impacts on $\amin$. Note that we have found that $\tau\sub{RGB}$ is not sensitive to stellar mass; for example, the periods during which $R_\star (t) > 0.5 R\sub{tip}$ in the RGB phase are 4.3~Myr, 9.7~Myr, and 5.2~Myr for 1.5$\Msun$, 2.0$\Msun$, and 3.0$\Msun$ stars with $Z_\star = 0.02$, respectively. {It is worth noting that equation~(\ref{eq:acrit}) agrees well with the numerical values of $\amin$ on the high-mass side in Fig.~\ref{fig:amin vs. aorb}; the differences are less than 10~\% for $M_{\star}\gtrsim 2.0 M_{\odot}$. On the low-mass side (i.e., $M_\star <2.0M_{\odot}$ ), the differences are bigger; for example, equation~(\ref{eq:acrit}) predicts that $\amin$ is bigger by about 25~\% (0.4AU) than its numerical value for $M_\star=1.5M_\odot$. This is because the stellar mass loss is relatively effective in pushing planets outwards in the low-mass cases.}

Next, as seen in Fig.~\ref{fig:amin vs. aorb}, there is a sharp transition in between 1.7~$\Msun$ and 2.1~$\Msun$: $\amin$ changes by approximately one order of magnitude in such a narrow mass range.  The transition corresponds to that of $R\sub{tip}$ that is related to the thermal state of the stellar helium core. A low-mass star of $\lesssim 2 \Msun$ has a degenerate helium core after the exhaustion of the central hydrogen and undergoes the helium flash, which causes significant expansion in its RGB phase; in contrast, an intermediate-mass star of $\gtrsim 2 \Msun$ soon starts the 3$\alpha$ burning at its center instead of the helium flash, so that $R\sub{tip}$ is relatively small.

\begin{figure}[tb]
  \begin{center} 
    \includegraphics[width=8cm,keepaspectratio]{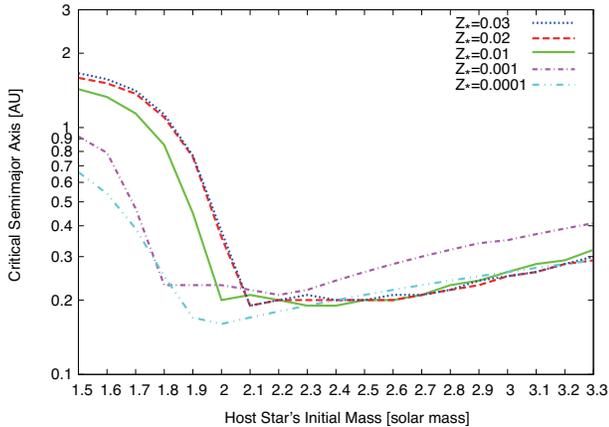}
    \caption{\small{The critical initial semimajor axis, $\amin$, inside which planets are engulfed by host stars before reaching the end of the helium-burning phase for five choices of stellar metallicity, $Z_\star$. Dotted, dashed, solid, dot-dashed, and double-dot-dashed lines represent the results for $Z_\star$ =0.03, 0.02, 0.01, 0.001, and 0.0001, respectively. In these calculations, the planet's mass is 1~$M\sub{J}$.} \label{fig:metallicity}}
  \end{center}
\end{figure}

Finally, the dependence of $\amin$ on $M_{\star, i}$ is qualitatively the same regardless of stellar metallicity ($Z_\star$), but $\amin$ decreases with decreasing $Z_\star$, as illustrated in Fig.~\ref{fig:metallicity}. This is also related to the dependence of $R\sub{tip}$ on $Z_\star$. Because of low opacity in the atmosphere and of low efficiency of the CNO cycle at the hydrogen-burning shell, metal-poor RGB stars expand less compared to metal-rich RGB stars.

\section{COMPARISON WITH OBSERVATION} \label{sec:discussion}

\begin{figure}[tb]
  \begin{center} 
     \includegraphics[width=8cm,keepaspectratio]{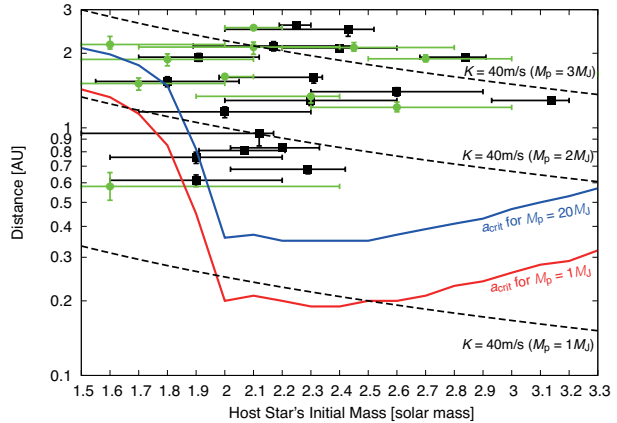}
    \caption{\small{Comparison between the theoretical survival limits and the semimajor axes of known planets around clump giants. The solid lines represent the critical initial semimajor axes, $\amin$, for stellar metallicity $Z_\star = 0.01$ and planetary mass $M_p =$ 1 and 20 $M\sub{J}$. The semimajor axes of the planets are represented by symbols with error bars: filled squares and circles correspond to values from the literature and those re-evaluated with MESA, respectively. The dashed lines represent radial-velocity semi-amplitude of 40 $\rm m\,s^{-1}$ for $M_p$ = 1, 2, and 3 $M\sub{J}$.}   \label{fig:observation}}
  \end{center} 
\end{figure}

{The critical semimajor axis, $\amin$, that we have calculated numerically is compared with measured semimajor axes of planets so far detected around GK clump giants in Figure~\ref{fig:observation}: }$\amin$ for $M_p =$ 1~$M\sub{J}$ and 20~$M\sub{J}$ and $Z_\star$ = 0.01 are represented by solid lines. The measured semimajor axes are represented by symbols with error bars---filled squares are data from the literature, while filled circles are values that we have re-evaluated using theoretical evolution curves generated with MESA. Published values of the stellar masses were evaluated based on other stellar codes \citep[][]{Girardi+00,Lejeune+01}.

Re-evaluation of stellar masses has been done just for self-consistency; that never means our values are more reasonable than the published ones. Because stars of interest in this study are located in a region crowded with evolutionary curves in the H-R diagram, it is natural that the estimation of stellar mass includes large uncertainty, which is beyond the scope of this study. Note that the three different codes, namely, MESA, \citet{Girardi+00}'s, and \cite{Lejeune+01}'s yield similar values of $\amin$, as found in section~\ref{sec:code}.  The evaluated values of stellar masses are listed on Table~\ref{tbl:stars}. As a result of stellar-mass re-evaluation, planetary masses and semimajor axes are also changed. These values are also listed on Table 1. In comparing observational data with theoretical curves, we have assumed that stars are located in HeB or post-HeB phases.

Figure~\ref{fig:observation} indicates two important things: First, one finds that almost all the symbols are above the curves of $\amin$. Although the mean values for a few planets are smaller than $\amin$, those error bars extend out to the curves of $\amin$. 
Second, however, one realizes that on the high-mass side (i.e., $M_\star > 2.1 \Msun$), planets exist far above $\amin$. There seems to be a gap between the theoretical survival-limit and the observed existence-limit of planets. Thus, our comparison with observation suggests that planet engulfment by host stars may not be the main reason for the lack of {\SPGP}, at least, for $M_\star > 2.1 \Msun$. {Note that we have to keep in mind that the number of planets may be still statistically insufficient.}

The presence of the gap is not due to observational biases. Clump giants typically show intrinsic radial-velocity variability of $\sigma$ = 10--20 m s$^{-1}$ (Sato et al. 2005). Given that we can only detect planets that impart radial-velocity semiamplitude of $K>40$ m s$^{-1}$ (i.e., 2$\sigma$) to their host stars, we indicate the detection limits by dashed lines for $M_p =$ 1, 2, and 3~$M\sub{J}$ in the figure. While the detectable limit for planets of $\leq$ 1 $M\sub{J}$ is below $\amin$, that for planets of $\geq$ 2 $M\sub{J}$ is significantly above $\amin$. Because the masses of the planets shown in the figure are all more than 2~$M\sub{J}$, the gap that we have found is not due to such observational biases.

\section{SENSITIVITY TO UNCERTAINTIES \label{sec:validity}}
\subsection{Tidal dissipation \label{sec:tidal model}}

For tidal dissipation, this study is based on the widespread idea of turbulent viscosity (eqs.~[\ref{eq:k}]-[\ref{eq:T}]).
It remains uncertain how to deal with fast tide when the planetary orbital period, $P$, is shorter than the eddy turnover timescale, $T$. For such a situation to be taken into account, the factor $f$ is added in equation~(\ref{eq:orbit}) to weaken the stellar tide. We have adopted the Kolmogorov scaling, namely, $f \propto (P/T)^2$, which supposes that  only small eddies whose intrinsic turnover times are shorter than the orbital period contribute to the tidal dissipation \citep[][]{Goldreich+Nicholson77}. However, the Kolmogorov picture is known to be too simplified for the turbulent viscosity. The dependence of the viscosity on $P/T$ is a matter of debate \citep[e.g.,][and references therein]{Zahn08,Penev+07}.
For example, recent hydrodynamic simulations by \citet{Penev+07} demonstrate that the dependence is close to linear (i.e., $f \propto P/T$), which happens to be consistent with that proposed by \citet{Zahn66} who assumed that the largest eddy dominates tidal dissipation.

\begin{figure}[tb]
  \begin{center} 
    \includegraphics[width=8cm,keepaspectratio]{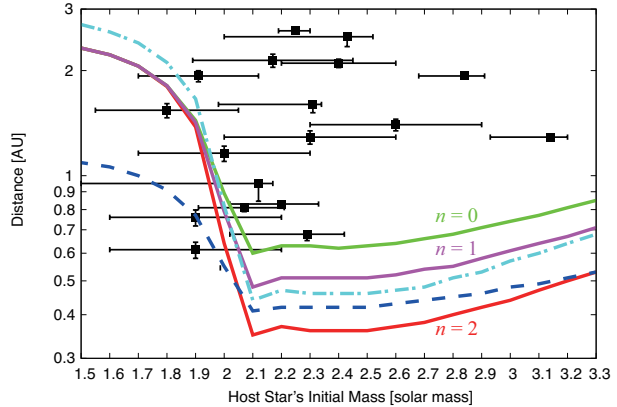}
    \caption{\small{{The impacts of uncertainties in turbulent viscosity and eccentricity on the critical initial semimajor
axes. The three solid lines show the results with different frequency-dependent turbulent viscosities, namely, the power index, $n$, of $f=\min[1,(P/2T)^n]$ being  0, 1, and 2. The dashed line corresponds to the case with constant quality factor $Q'_{\star}=10^5$ (see text for more detail). The dash-dotted line corresponds to the case with finite eccentricity of 0.25 (and $n$ = 2). In those calculations, $Z_\star =0.02$ and $M_p=20 M\sub{J}$. The symbols with error bars represent the semimajor axes of the planets detected around GK clump giants.}} \label{fig:tidal model}}
  \end{center} 
\end{figure}

Figure~\ref{fig:tidal model} shows $\amin$ calculated for three different values of the power index $n =$ 0, 1 and 2 (solid lines) for $f=\min[1,(P/T)^n]$. It also plots the measured semimajor axes of the planets of interest.  In calculating $\amin$, we assumed the planet's mass is 20~$M_J$, approximately the largest value of the minimum masses of planets presented in Fig.~\ref{fig:tidal model}.

As $n$ increases, the tidal dissipation becomes less efficient, when $P < T$. 
In other words, decreasing $n$ moves the curve of $\amin$ upwards in Fig.~\ref{fig:tidal model}. 
As seen in the figure, $\amin$ for $n = 1$ is larger by about 50\% than that for $n=2$ on the high-mass side. Note that $\amin$ hardly differs on the low-mass side, because $\amin$ is so large that $P$ is almost always larger than $T$, namely, $f = 1$.
In an extreme case of $n = 0$ (i.e., $f = 1$), $\amin$ is about twice as large as that for $n = 2$ on the high-mass side.
Even in this case, decreasing $n$ does not fill the gap.

Finally we comment on the $Q$-prescription for stellar tide. Equation~(\ref{eq:orbit}) can be written in the form by the use of the quality factor, $Q^\prime_\star$ \citep[e.g.,][]{Goldreich+Soter66},
   \begin{equation}
     \frac{1}{a} \frac{\mathrm{d}a}{\mathrm{d}t} = 
        -\frac{9}{2} \left( \frac{G}{M_{\star}}\right) ^{1/2} \frac{M_p}{Q'_\star}
		       \frac{R_\star ^5}{a^{13/2}}
	-\frac{\dot{M}_\star}{M_\star}.
   \label{eq:CPL}
   \end{equation}
In this study, $Q^\prime_\star$ varies with time, while $Q^\prime_\star$ is often assumed to be constant ($= 10^5$-$10^8$ for MS stars in the literature). 
\citet{Nordhaus+10} studied tidal engulfment mainly by AGB stars in both cases of variable $Q'_{\star}$ ($f=1$) and constant $Q'_{\star}$ (= $10^5$).
They found that the tidal decay rate in the former case is about 4 orders of magnitude higher than that in the latter case. 
In many cases of this study, $f = (P/2T)^2 \ll 1$, instead of $f =1$. Thus, it would be worth checking the outcome for constant $Q^\prime_\star$ in our high-mass cases. The dashed line in Fig.~\ref{fig:tidal model} shows $\amin$ calculated by equation~(\ref{eq:CPL}) with $Q^\prime_\star = 1 \times 10^{5}$. 
As seen in the figure, the tidal decay with $Q^\prime_\star = 1 \times 10^5$ yields similar values of $\amin$ that we derive with variable $Q^\prime_\star$ on the high-mass side. 
This means stellar tide with high $Q^\prime_\star$ does not fill the gap neither.

\subsection{Stellar Evolution \label{sec:code}}

There are several different stellar-evolution models in the literature, as mentioned in section~\ref{sec:method}. We discuss the differences in terms of the survival limit, and clarify the cause of the relatively big difference in $\amin$ between this study and \citet{Villaver+Livio09}.

\begin{figure}[tb]
  \begin{center} 
    \includegraphics[width=8cm,keepaspectratio]{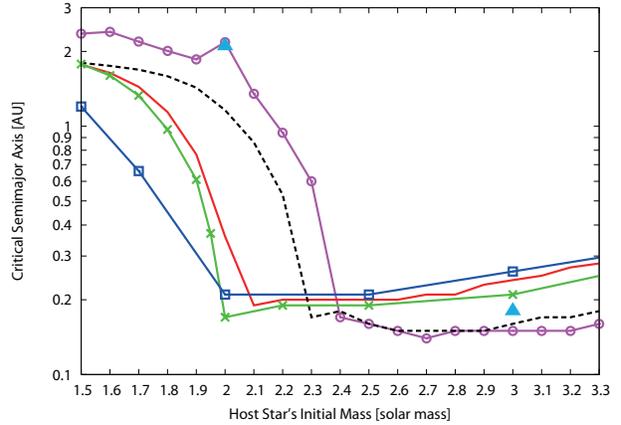}
    \caption{\small{The critical initial semi-major axes  as a function of stellar mass calculated with different four stellar-evolution codes: The solid and dashed lines are from MESA~v2258 \citep{Paxton+10} with and without overshooting, respectively; the lines with crosses, with open squares, and with open circles are from \citet{Girardi+00}'s, \citet{Lejeune+01}'s and \citet{Suda+07}'s codes, respectively; the filled triangles are from \citet{Villaver+Livio09}. 
The critical initial semi-major axes (except the triangles) have been calculated until the host star's RGB-tip under the assumptions that $M\sub{env} = M_\star$ and $R\sub{env} = 0$ in the calculations of stellar tide (see eqs~[\ref{eq:k}] and [\ref{eq:T}]) and that $\dot{M}_\star = 0$ in equation~(\ref{eq:orbit}). As for calculations with \citet{Girardi+00}' code, we have assumed that $M_{\star}$ is constant with time because values of $M_{\star}$ as a function of time are unavailable.} \label{fig:acrit-codes}}
  \end{center}
\end{figure}

In Fig.~\ref{fig:acrit-codes}, the solid and dashed lines (without symbol) represent results that we have obtained using MESA with and without overshooting, respectively. The line with open circles is the result that we have calculated using \citet{Suda+07}'s code. The two triangles correspond to the values of $\amin$ from \citet{Villaver+Livio09}. Those two codes do not include overshooting. The lines with crosses and squares are the results calculated with grid data from \citet{Girardi+00} and \citet{Lejeune+01}, respectively, which include overshooting; those code have been used to estimate the mass of planet-host clump giants.

When deriving $\amin$ in Fig.~\ref{fig:acrit-codes}, except for those from \citet{Villaver+Livio09}, we have assumed that $M\sub{env} = M_\star$ and $R\sub{env} = 0$ in calculating quantities relevant to stellar tide, $k$ and $T$ (see eqs.~[\ref{eq:k}] and [\ref{eq:T}]), because no detailed information about the stellar internal structure is available. {Such simplification certainly results in inaccurate evaluation of tidal dissipation (i.e., $k/T$), but the tidal dissipation factor was found to play a minor role in $\amin$ in section~\ref{sec: survival limit}; the most important factor is the stellar radius. This means that such simplification as to stellar structure does not affect our conclusions in this section. }

We have figured out that inclusion of overshooting probably causes the difference in $\amin$ between our calculations (solid line without symbol) and \citet{Villaver+Livio09}'s (triangles). Qualitatively, overshooting plays a role in mixing hydrogen and helium, so that stellar models with overshooting contain bigger helium cores compared to those without overshooting. This results in higher central temperature, which prevents the helium core from being degenerated. This means that overshooting is effective in moving the sharp transition from the high-$\amin$ to the low-$\amin$ domains leftwards in Fig.~\ref{fig:acrit-codes}. That is why we have obtained a value of $\amin$ quite different from that by \citet{Villaver+Livio09} for $M_{\star} = 2 \Msun$. 

In any case, the drastic decrease in $a\sub{crit}$ due to avoidance of helium flash does occur, independent of simulation code, although the threshold mass somewhat depends on code. {Difference in code never changes our conclusion concerning the survival limit.}

\subsection{Eccentricity \label{sec:e}}
{Since we are interested in semimajor axes of planets, we have integrated the equation for change in semimajor axis (eq.~[\ref{eq:orbit}]), ignoring eccentricity, following previous studies \citep[e.g.,][]{Villaver+Livio09,Nordhaus+10}. However, the planets detected around clump giants have finite eccentricities. Here we make a simple check of the impact of the finite eccentricity on the critical semimajor axis.}

{The dash-dotted line in Fig.~\ref{fig:tidal model} shows $\amin$ for $e = 0.25$ that we have calculated by integrating the equation \citep[e.g., see][]{Hut81}, 
     \begin{eqnarray}
     \frac{1}{a} \frac{\mathrm{d}a}{\mathrm{d}t} &=& 
        -6 \frac{k}{T} \frac{M_p}{M_\star} 
	               \left(1+\frac{M_p}{M_\star}\right)
		       \left(\frac{R_\star}{a}\right)^8 
			\frac{f_1(e^2)}{(1-e^2)^{15/2}}  \nonumber \\
        & &	-\frac{\dot{M}_\star}{M_\star},
   \label{eq:orbit-e}
  \end{eqnarray}   
where
   \begin{equation}
		f_1(e^2)=1+\frac{31}{2}e^2+\frac{255}{8}e^4+\frac{185}{16}e^6+\frac{25}{64}e^8.
	\label{eq:f1}
	\end{equation}
The likelihood values of eccentricities of inner planets ($a<1.9$~AU) are smaller than $0.25$. 
In this figure, it is demonstrated that incorporating the finite value of $e = 0.25$ does not affect our conclusion about the presence of the gap. Note that the evolution of  $e$ itself coupled with that of host stars' rotation is an interesting problem, which is our future work. 
}

\section{STELLAR-MASS-DEPENDENT PLANET FORMATION \label{sec:primordial}}

A possibility other than planet engulfment is that some stellar-mass-dependent processes hinder the formation of {\SPGP} around 1.5-3$\Msun$ stars.  
For example, ultraviolet measurements of young stars indicate that the stellar accretion rate ($\dot{M}_\star$) increases rather rapidly with the stellar mass, namely,  $\dot{M}_\star \propto M_\star^{2}$ \citep{Calvet+04,Muzerolle+05}. 
This implies that lifetimes of protoplanetary disks of heavier stars are shorter. 
Based on this fact, \citet{Burkert+Ida07} supposed that protoplanetary disks could dissipate before giant planets migrate to the vicinity of the central stars in the case of F stars  ($1.2$-$1.5\Msun$). Then, they demonstrated that the idea worked to account for the observed period valley between the hot-Jupiter and cool-Jupiter classes. Applying the idea to heavier stars, \citet{Currie09} proposed that {\SPGP} are rarely formed around $\sim$1.5--3.0 $\Msun$ stars.

However, the primordial origin of the paucity of {\SPGP} remains a matter of debate. 
The negative dependence of disk lifetime on stellar mass has not yet confirmed theoretically. 
It is also controversial whether planetary migration occurs in disks for which such significant stellar accretion is observed. Furthermore, there are several competing factors that affect the frequency of occurrence of {\SPGP} around $\sim 1.5$-$3$~$\Msun$ stars. More distant snowlines due to high luminosity of such stars also tend to reduce the frequency (Ida \& Lin 2005). On the other hand, more massive disks and shorter Kepler periods at given semimajor axes form giant planets faster, which results in higher production rate of {\SPGP} and suggests that detection frequency of giant planets may be higher around high-mass stars \citep[][]{Omiya+09}. Further detailed observational data on SPGPs will constrain stellar mass dependence of planet formation.

\section{SUMMARY AND CONCLUSIONS \label{sec:conclusion}}

In this study we have derived the survival limit beyond which planets are engulfed by their host stars in the RGB and HeB phases, by numerically simulating both the tidal evolution of planetary orbits and the evolution of the host stars. Then, we have made a detailed comparison of the survival limit with observed semimajor axes of planets detected so far around GK clump giants. The comparison demonstrates that (1) almost all the planets are orbiting exterior to our theoretical survival limit and (2) the planets around stars with relatively large masses of $\gtrsim 2.5 \Msun$ are far outside the survival limit. Those facts suggest that planet engulfment by host stars may
actually happened, but it may never be the main reason for the observed lack of short-period planets around clump giants.  
Because the number of planets is statistically insufficient in that high-mass range, we have to say that our findings are not definite. The problem we addressed have a great influence on our understanding of planet formation.
Therefore, further surveys for planets around clump giants are highly encouraged.

\acknowledgments

We would like to express our gratitude to the following persons: Bill Paxton and Aaron Dotter kindly helped us install and use the stellar-evolution code MESA and modified it upon our request. Masayuki Y. Fujimoto and Takuma Suda gave useful comments about stellar evolution. We had fruitful discussion on this study with Yasunori Hori and Taishi Nakamoto. We appreciate the anonymous referee's critical and constructive comments that helped us to improve this paper.  This work is supported partly by JSPS Grant-in-Aid for Scientific Research (A) (No. 20244013).


\clearpage
\begin{table}
 \begin{center}
 \caption{Properties of stars and planets re-evaluated \label{tbl:stars}}
  \begin{tabular}{lcccccc}
    \tableline\tableline
	Stellar Name & $M_\star$ [$\Msun$]\tablenotemark{a} & $M_\star$ [$\Msun$] & $M_p\,\sin{i}$ [$M\sub{J}$] & $a$ [AU] &Ref.\tablenotemark{b}\\
    \tableline \\
    11 Com\tablenotemark{c}      &   3.14$+0.06\atop-0.21$     & 2.6$+0.4\atop-0.3$        & 17.1$+1.7\atop-1.3$ & 1.21$+0.06\atop-0.05$ & (1),(2) \\ 
	11 UMi\tablenotemark{d}      & 1.8$+0.25\atop-0.25$       & 1.7$+0.3\atop-0.3$        & 10.8$+1.2\atop-1.3$ & 1.51$+0.09\atop-0.09$ & (3)\\
	14 And\tablenotemark{c}	&  2.20$+0.13\atop-0.18$	& 1.2$+0.2\atop-0.3$        & 3.20$+0.4\atop-0.6$ & 0.68$+0.03\atop-0.06$ & (4),(2)\\
	18 Del\tablenotemark{c}      &  2.25$+0.06\atop-0.05$    & 2.1$+0.1\atop-0.1$        & 9.84$+0.3\atop-0.3$ & 2.54$+0.04\atop-0.04$ & (5),(2)\\
	81 Cet\tablenotemark{c}      &   2.43$+0.09\atop-0.43$      & 1.6$+0.4\atop-0.2$        & 4.01$+0.6\atop-0.3$ & 2.17$+0.17\atop-0.09$ & (4),(2)\\
	HD 102272\tablenotemark{d}	&   1.9$+0.3\atop-0.3$   & 1.6$+0.8\atop-0.3$        & 5.26$+1.6\atop-1.2$ & 0.58$+0.08\atop-0.07$ & (6)\\
	HD 104985\tablenotemark{c}	&  2.12$+0.05\atop-0.63$    & NA                                 &                            &		  & (5),(2)\\
	HD 110014\tablenotemark{d}	&   2.17$+0.28\atop-0.28$   & 2.1$+0.3\atop-0.4$        & 10.9$+1.0\atop-1.4$ & 2.12$+0.09\atop-0.15$ & (7)\\
	HD 119445\tablenotemark{e}	&   3.9$+0.4\atop-0.4$    & 3.5$+0.3\atop-0.2$        & 35.0$+2.0\atop-1.4$ & 1.65$+0.05\atop-0.03$ & (8)\\
	HD 11977\tablenotemark{d}    &  1.91$+0.21\atop-0.21$    & 1.8$+0.3\atop-0.3$        & 6.29$+0.7\atop-0.7$ & 1.89$+0.10\atop-0.11$ & (9)\\
	HD 145457\tablenotemark{e}	&   1.9$+0.3\atop-0.3$     & 1.0$+0.4\atop-0.1$        & 1.89$+0.5\atop-0.1$ & 0.61$+0.08\atop-0.02$ & (10)\\
	HD 17092\tablenotemark{d}    &  2.3$+0.3\atop-0.3$   & 1.55$+0.65\atop-0.25$  & 3.53$+0.9\atop-0.4$ & 1.13$+0.14\atop-0.06$ & (11)\\
	HD 173416\tablenotemark{e}	&  2.0$+0.3\atop-0.3$  & 1.3$+0.4\atop-0.4$        & 2.03$+0.4\atop-0.4$ & 1.00$+0.10\atop-0.11$ & (12)\\
	HD 180314\tablenotemark{e}	&   2.6$+0.3\atop-0.3$    & 2.3$+0.1\atop-0.4$        & 20.3$+0.6\atop-2.4$ & 1.34$+0.02\atop-0.08$ & (10)\\
	HD 62509\tablenotemark{c}     &  2.31$+0.03\atop-0.33$   & 2.0$+0.1\atop-0.0$        & 2.63$+0.1\atop-0.0$ & 1.61$+0.03\atop-0.00$ & (13),(2)\\
	HD 81688\tablenotemark{c}   &   2.07$+0.14\atop-0.16$    & 1.1$+0.3\atop-0.2$        & 1.77$+0.3\atop-0.2$ & 0.66$+0.05\atop-0.05$ & (5),(2)\\
	$\epsilon$ Tau\tablenotemark{c}	&  2.84$+0.07\atop-0.16$   & 2.7$+0.3\atop-0.2$        & 7.34$+0.5\atop-0.4$ & 1.90$+0.07\atop-0.05$ & (14),(2)\\
	$\xi$ Aql\tablenotemark{c}	&   2.29$+0.27\atop-0.13$    & 1.4$+0.2\atop-0.2$       & 2.02$+0.2\atop-0.2$ & 0.58$+0.02\atop-0.03$ & (5),(2)\\
	NGC 2423\tablenotemark{f}    &  2.4$+0.2\atop-0.2$   & 2.45$+0.35\atop-0.25$ & 10.7$+1.0\atop-0.7$ & 2.11$+0.10\atop-0.07$ & (15)\\
	NGC 4349\tablenotemark{f}    &  3.9$+0.3\atop-0.3$  & 3.9$+0.5\atop-0.5$       & 19.8$+1.7\atop-1.7$ & 2.38$+0.10\atop-0.11$ & (15)\\    
    \tableline
  \end{tabular}
 \tablecomments{$M_\star$,  $M_p \sin{i}$, and $a$ are the stellar mass, the planetary minimum mass, and the semimajor axis, respectively.  NA means that under the assumption that the star is on its HeB or post-HeB, no theoretical track matches its luminosity and effective temperature in the H-R diagram.  The samples above listed are stars with published values of masses $M_{\star} \geq 1.6 M_{\odot}$ and radii $R_{\star} \simeq 10-20 R_{\odot} $.  When re-evaluating, we have assumed that the solar effective temperature is 5777~K and the solar metallicity is 0.02. }
 \tablenotetext{a}{Stellar masses, $M_{\star}$, in the literature. }
 \tablenotetext{b}{(1) \citet{Liu+08}; (2) \citet{Takeda+08}; (3) \citet{Dollinger+09}; (4) \citet{Sato+08b}; (5) \citet{Sato+08}; (6) \citet{Niedzielski+09a}; (7) \citet{de Medeiros+09}; (8) \citet{Omiya+09}; (9) \citet{Setiawan+05}; (10) \citet{Sato+10}; (11) \citet{Niedzielski+07}; (12) \citet{Liu+09}; (13) \citet{Hatzes+06}; (14) \citet{Sato+07}; (15) \citet{Lovis+Mayor07}}
  \tablenotetext{c}{Values of the effective temperature, $T\sub{eff}$, and the luminosity, $L_{\star}$ and their errors have been taken from the references.}
  \tablenotetext{d}{Values and their errors of $L_{\star}$ have been derived form those of $T\sub{eff}$ and the stellar radius, $R_\star$. }
  \tablenotetext{e}{Errors of $L_{\star}$ have been derived from those of $T\sub{eff}$ and $R_\star$. }
  \tablenotetext{f}{Stellar masses have been determined so that the stellar ages match the observed cluster age, the stars being assumed to be on the HeB.}
 \end{center}
\end{table}

\end{document}